\documentclass[a4paper,amssymb,amsmath,twocolumn,longbibliography,accepted=2021-10-01]{quantumarticle}
\pdfoutput=1
\usepackage{graphicx}
\usepackage{hyperref}
\usepackage{placeins}
\usepackage{xcolor}
\usepackage{listings}
\usepackage[normalem]{ulem}
\usepackage[utf8]{inputenc}
\usepackage{color}
\usepackage{float}
\usepackage{simplewick}

\usepackage{epstopdf}

\begin{document}
	
\title{Quantum-classical correspondence of a system of interacting bosons in a triple-well potential}

\author{E. R. Castro}
\email[]{ercmlv@gmail.com}
\affiliation{Instituto de F\'{\i}sica da UFRGS Av. Bento Gon\c{c}alves 9500, Porto Alegre, RS, Brazil} 
\affiliation{Centro Brasileiro de Pesquisas F\'{\i}sicas/MCTI,
	22290-180, Rio de Janeiro, RJ, Brazil} 

\author{Jorge Ch\'avez-Carlos}
\email[]{jorge.chavez@ciencias.unam.mx}
\affiliation{Instituto de Ciencias F\'{\i}sicas, Universidad Nacional Aut\'onoma de M\'exico,
	Cuernavaca, Morelos 62210, M\'exico}

\author{I. Roditi}
\email[]{roditi@cbpf.br}
\affiliation{Centro Brasileiro de Pesquisas F\'{\i}sicas/MCTI,
	22290-180, Rio de Janeiro, RJ, Brazil}

\author{Lea F. Santos}
\email[]{lsantos2@yu.edu}
\affiliation{Department of Physics, Yeshiva University, New York, New York 10016, USA}

\author{Jorge G. Hirsch}
\email[]{hirsch@nucleares.unam.mx}
\affiliation{Instituto de Ciencias Nucleares, Universidad Nacional Aut\'onoma de M\'exico,
	Apdo. Postal 70-543, C.P. 04510 Cd. Mx., M\'exico}

\maketitle
\begin{abstract}
	We study the quantum-classical correspondence of an experimentally accessible system of interacting bosons in a tilted triple-well potential. With the semiclassical analysis, we get a better understanding of the different phases of the quantum system and how they could be used for quantum information science. In the integrable limits, our analysis of the stationary points of the semiclassical Hamiltonian reveals critical points associated with second-order quantum phase transitions. In the nonintegrable domain, the system exhibits crossovers. Depending on the parameters and quantities, the quantum-classical correspondence holds for very few bosons. In some parameter regions, the ground state is robust (highly sensitive) to changes in the interaction strength (tilt amplitude), which may be of use for quantum information protocols (quantum sensing).
\end{abstract}

\section{Introduction}

Studies of the quantum-classical correspondence provide insights into the properties of both the quantum system and its classical counterpart. Level statistics as in full random matrices~\cite{MehtaBook}, for example, is a quantum signature of classical chaos~\cite{Casati1980, Bohigas1986}. In the other direction, classical chaos and instability are related with the exponential growth of the out-of-time-ordered correlator~\cite{Rozenbaum2017,Hashimoto2020}, and unstable periodic orbits explain the phenomenon of quantum scarring~\cite{Heller1984,Pilatowsky2021}. 
In this work, we explore the quantum-classical correspondence for yet another goal, that of locating the quantum phase transition points of a system of interacting bosons in a triple-well potential.

Previous semiclassical analyses of three coupled Bose-Einstein condensates have revealed a dynamical transition from self trapping to delocalization~\cite{Nemoto2000,Liu2007}. The quantum dynamics in triple well traps have since been extensively investigated~\cite{Buonsante2009,Richaud2018}. When quantum gases, such as chromium or dysprosium, are loaded into triple well potentials~\cite{Lahaye2010}, dipolar interactions need to be taken into account~\cite{Lahaye2010} and they lead to various ground-state phases~\cite{Lahaye2010,DellAnna2013}. An integrable version of this dipolar model in one dimension, solvable with the algebraic Bethe ansatz, was derived in~\cite{Ymai2017}, and by tilting the potential, this model can be brought to the chaotic domain~\cite{Wittmann2018}. The tilt is an additional control parameter that expands the versatility of the model and allows for its possible application as an atomtronic switching device~\cite{Wittmann2018} and as a generator of entangled states~\cite{Tonel2020}. This is the system that we analyze here, both in its integrable and nonintegrable regimes.

With the semiclassical Hamiltonian of our triple-well system, we find the stationary points of the classical dynamics and use them to identify the critical points of quantum phase transitions. We show that there are two integrable limits that exhibit second-order quantum phase transitions. One critical point is accessed by varying the interaction strength between the bosons with respect to the tilt of the potential, while the tunneling amplitude between the wells is zero, and the other point is found by changing the interaction strength with respect to the tunneling amplitude, while the tilt is zero. The different phases are characterized by different values of the occupation numbers of the wells, which serve as good order parameters. In the nonintegrable domain, where the three Hamiltonian parameters are nonzero, the system exhibits crossovers.  

We find that the nonintegrable model presents two interesting features with potential applications. In the region of attractive interaction, the occupation numbers of the wells at the edges of the chain are highly sensitive to the amplitude of the tilt, which could be explored for developing quantum sensors. In the opposite region of repulsive interaction, the ground state is protected against changes of the interaction strength and of the tilt amplitude around zero, a feature that might be advantageous for certain quantum information protocols.

In addition, we analyze how the quantum-classical correspondence for the lowest energy state depends on the total number of bosons. This point is related with the question of how many particles are needed for a system to reveal many-body features~\cite{Blume2010,FogartyARXIV}, a subject of current experimental interest~\cite{Serwane2011,Wenz2013}. In the absence of interaction, the agreement is exact for a single boson, since the system is in the semiclassical limit. In the presence of tunneling and interaction, and for a finite range of values of the tilt, the coincidence between the quantum and semiclassical results can also hold for a single particle depending on the quantity and whether the interaction is attractive or repulsive. 

The paper is organized as follows. Section~II describes the Hamiltonian and the stationary points. Section~III is dedicated to the analysis of the three integrable limits of the model and Sec.~IV to the analysis of the nonintegrable regime. Our conclusions are presented in Sec.~V.

\section{Quantum and classical Hamiltonians}

In this section, after describing the quantum Hamiltonian, we explain its semiclassical limit and how to determine the stationary points.

\subsection{Quantum Hamiltonian}

\begin{figure}[ht]
	\centering
	\includegraphics[width=1.\linewidth]{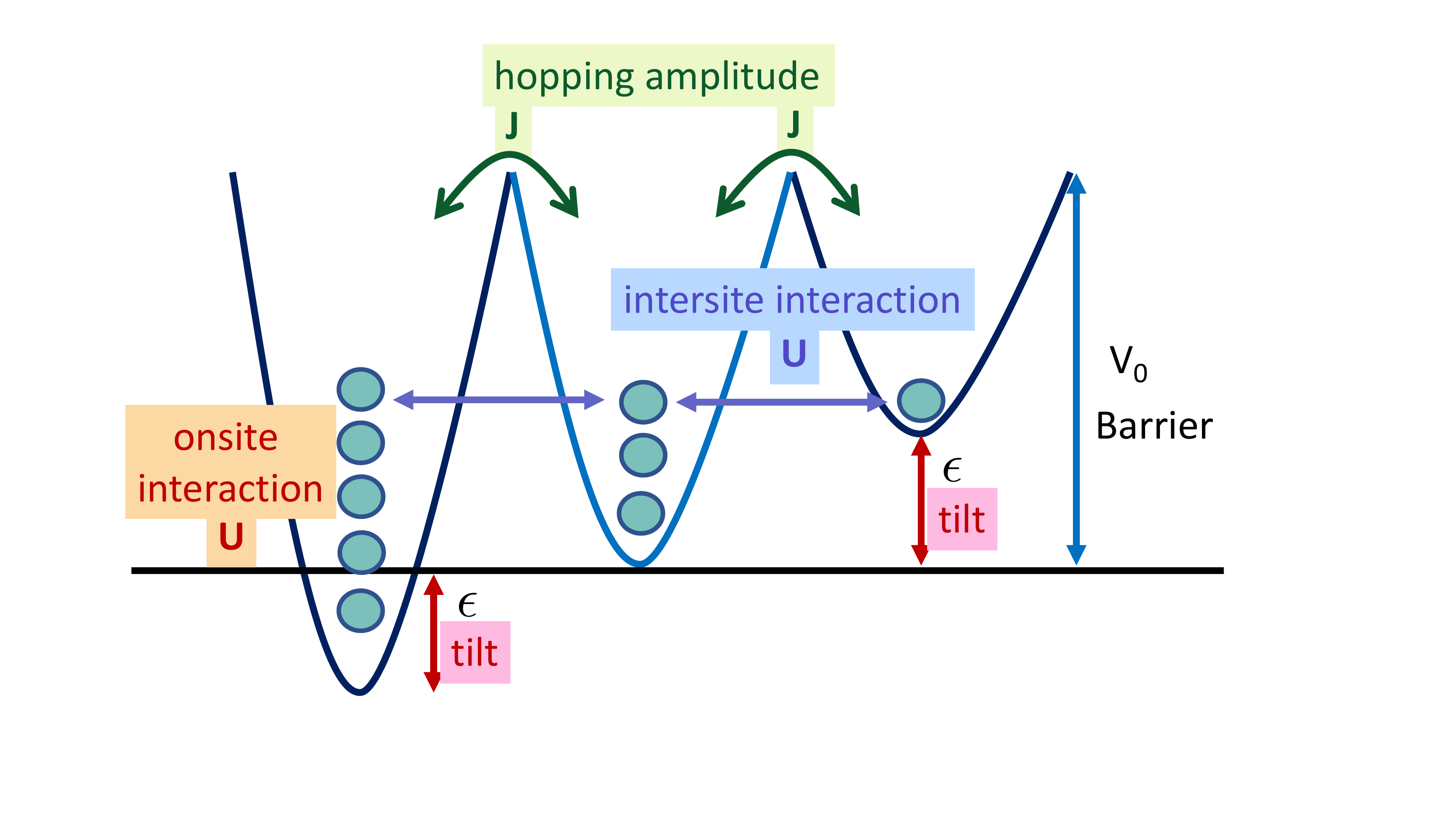}
	\caption{Schematic representation of the three-well system. }
		\label{fig00}
	\end{figure}

We consider $N$ bosons in an aligned three-well potential [see Fig.~\ref{fig00}]. The quantum  Hamiltonian is given by
\begin{align}
\label{QH}
	\hat{H} =& \frac{U}{N}\left(\hat{N}_1-\hat{N}_2+\hat{N}_3\right)^2 + \epsilon\left(\hat{N}_3-\hat{N}_1\right) \nonumber \\ &+\frac{J}{\sqrt{2}}\left(\hat{a}_1^\dagger \hat{a}_2 + \hat{a}_2^\dagger \hat{a}_1\right)+\frac{J}{\sqrt{2}}\left(\hat{a}_2^\dagger \hat{a}_3 + \hat{a}_3^\dagger \hat{a}_2\right),
\end{align}
where $\hat{N}_k=\hat{a}_k^\dagger \hat{a}_k$ is the number operator of the well $k$, $\hat{a}_k$ ($\hat{a}_k^\dagger$) is the annihilation (creation) operator, $U$ represents both the onsite interaction strength and the strength of the interactions between wells, and it is rescaled by $N$, $J$ is the tunneling amplitude between wells, and $\epsilon$ is the tilt. The Hamiltonian is invariant under the interchange of wells 1 and 3 when $\epsilon=0$, and it conserves the total number of bosons, $N=N_1+N_2+N_3$, having dimension $D=(N+2)!/(2!N!)$. Our analysis is conveniently done in the Fock basis representation, $|N_1, N_2, N_3 \rangle$ and we use exact diagonalization~\cite{fn1}. 

The Hamiltonian presents three integrable limits: \\(A) $\pmb{U = 0}$, $\epsilon \neq 0 $, $J \neq 0$; \\(B) $\pmb{J = 0}$, $U \neq 0 $, $\epsilon \neq 0$; \\(C) $\pmb{\epsilon = 0}$, $U \neq 0 $, $J \neq 0$. \\As we show next, (B) and (C) exhibit quantum phase transitions at the critical points $U=\epsilon/4$ and $U=-J/2$, respectively. The model becomes nonintegrable when $J$, $U$ and $\epsilon$ are nonzero.

\subsection{Classical Hamiltonian and Stationary Points}

The classical Hamiltonian is obtained using coherent states~\cite{Hepp1974}, $|\alpha\rangle=|\alpha_1,\alpha_2,\alpha_3 \rangle$, where ${\alpha_k=\sqrt{N_k}\exp(i\phi_k)}$. It leads to
\begin{align}
\label{CH1}
	\mathcal{H}_{\text{cl}} =& \frac{\langle \alpha |\hat{H} |\alpha \rangle}{N}   \\ = &\frac{U}{N}\left(N_1-N_2+N_3\right)^2 + \epsilon\left(N_3-N_1\right) \nonumber + 
	J \sqrt{2} \\
	& \left[\sqrt{N_1N_2}\cos(\phi_1-\phi_2)+\sqrt{N_2N_3}\cos(\phi_2-\phi_3)\right] . \nonumber 
\end{align}

To simplify our analysis, we introduce a convenient set of classical variables, $\rho_k=\sqrt{N_k/N}$. In addition, since the quantum Hamiltonian conserves the total number of particles, we also impose the constraint $N=N_1+N_2+N_3$. The classical Hamiltonian now becomes 
\begin{align}
\label{CH2}
\bar{\mathcal{H}}_{\text{cl}} =&	\frac{\mathcal{H_{\text{cl}}}}{N} =
\, U\left(\rho_1^2-\rho_2^2+\rho_3^2\right)^2 + \epsilon\left(\rho_3^2-\rho_1^2\right) \nonumber \\ 
+&J \sqrt{2} \left[\rho_1\rho_2\cos(\phi_1-\phi_2)+\rho_2\rho_3\cos(\phi_2-\phi_3)\right] \nonumber \\
+&\lambda \left(1-\rho_1^2-\rho_2^2-\rho_3^2\right) ,
\end{align}
where $\lambda$ is the Lagrange multiplier associated with the constraint. 

The dynamical variables of the classical system are the $\rho_k$'s and the phase differences $\phi_{k,k+1}=\phi_k-\phi_{k+1}$. To find the stationary points, we first compute the partial derivatives in the phase-difference variables,
\begin{equation}
\label{StabPhi}
\frac{\partial \bar{\mathcal{H}}_{\text{cl}}}{\partial \phi_{12}}=\sin\phi_{12}=0 \hspace{0.7 cm } \frac{\partial \bar{\mathcal{H}}_{\text{cl}}}{\partial \phi_{23}}=\sin\phi_{23}=0 ,
\end{equation}
which gives $\phi_{12}=n\pi$ and $\phi_{23}=m\pi$, where $n,m$ are integers. 

Since the cosines of the phase differences are limited to $\pm 1$, they can be absorbed as a phase to $\rho_k$. This means that to obtain the stationary points, we can use the simplified Hamiltonian,
 \begin{align}
 \label{CH1EqC}
\bar{\mathcal{H}}_{\mathrm{eq}} =& U\left(\rho_1^2-\rho_2^2+\rho_3^2\right)^2 + \epsilon\left(\rho_3^2-\rho_1^2\right) \nonumber \\ 
+ &J \sqrt{2} \left(\rho_1\rho_2+\rho_2\rho_3\right) \nonumber \\
+&\lambda\left(1-\rho_1^2-\rho_2^2-\rho_3^2\right),
\end{align}   
and solve the following fours equations in the variables $(\rho_1,\rho_2,\rho_3,\lambda)$,
\begin{equation}
\label{StabNiC}
\frac{\partial \bar{\mathcal{H}}_{\mathrm{eq}}}{\partial \rho_1}=\frac{\partial \bar{\mathcal{H}}_{\mathrm{eq}}}{\partial \rho_2}=\frac{\partial \bar{\mathcal{H}}_{\mathrm{eq}}}{\partial \rho_3}=\frac{\partial \bar{\mathcal{H}}_{\mathrm{eq}}}{\partial \lambda}=0.
\end{equation}
The values of the phase differences are then simply inferred from the sign of $\rho_k \rho_{k+1}$, a positive (negative) value for $\rho_k \rho_{k+1}$ implies that $\phi_{k,k+1}$ is zero ($\phi_{k,k+1}=\pi$).

\begin{table*}
	\begin{center}
		\caption{{\bf Case $\pmb{U=0}$}: Exact expressions for the energies and variables ${(N_1,N_2,N_3, \phi_{12}, \phi_{23})}$ of the three stationary points.}
		\begin{tabular}{|c|c|c|c|c|c|} \hline
			Point & Energy $E/N$  & $N_1/N$ & $N_2/N$ & $N_3/N$ & $(\phi_{12},\phi_{23})$ \\ 
			\hline 
		\rule{0pt}{\dimexpr.7\normalbaselineskip+2.5mm}
			$x_1$ & $0$ & $\dfrac{J^2}{2(\epsilon^2+J^2)}$ & $\dfrac{\epsilon^2}{(\epsilon^2+J^2)}$ & $\dfrac{J^2}{2(\epsilon^2+J^2)}$ & $(0,\pi)$  \\ 
			\hline 
			$x_2$ & $-\dfrac{|\epsilon|\sqrt{\epsilon^2+J^2}}{\epsilon}$ & $\dfrac{[ \epsilon^2+J^2+\sqrt{\epsilon^2\left(\epsilon^2+J^2\right)  } ]^2}{4\left(\epsilon^2+J^2\right)^2}$ & $\dfrac{J^2}{2(\epsilon^2+J^2)}$ &$\dfrac{ [  \epsilon^2+J^2-\sqrt{\epsilon^2\left(\epsilon^2+J^2\right)  } ]^2}{4\left(\epsilon^2+J^2\right)^2}$ & $(\pi,\pi)$ \\  
			\hline 
			$x_3$ & $\dfrac{|\epsilon|\sqrt{\epsilon^2+J^2}}{\epsilon}$ & $\dfrac{ [   \epsilon^2+J^2-\sqrt{\epsilon^2\left(\epsilon^2+J^2\right)  } ]^2 }{4\left(\epsilon^2+J^2\right)^2}$ & $\dfrac{J^2}{2(\epsilon^2+J^2)}$ &$\dfrac{[  \epsilon^2+J^2+\sqrt{\epsilon^2\left(\epsilon^2+J^2\right) } ]^2}{4\left(\epsilon^2+J^2\right)^2}$ & $(0,0)$ \\ 
			\hline
		\end{tabular}  
		\label{tab2}
	\end{center}
\end{table*}

\section{Integrable limits}

The analysis of the integrable points provides the minimum energies of the semiclassical limit exactly and serves as a preparation for the study of the nonintegrable regime. 

\subsection{Case $\pmb{U=0}$}

\begin{figure}[htb]
	\includegraphics[width=1.\linewidth]{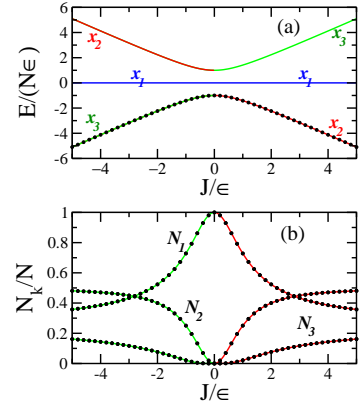}
	\caption{{\bf Case $\pmb{U=0}$}: Energies of the 3 stationary points (a) and occupation numbers $N_1$, $N_2$, and $N_3$ for the minimum energy state (b); $J>0$ and $N=20$.  Solid lines are for the semiclassical results and circles for the quantum ground state. 
	}
	\label{FigU0}
\end{figure}

In the absence of interaction, $U=0$, there is no phase transition. This regime is referred to as Rabi in the case of two wells and no tilt~\cite{Leggett2001} and the term has been borrowed also for the tilted triple-well potential~\cite{Wittmann2018}.
The solution of Eq.~(\ref{StabNiC}) reveals three stationary points, $x_1$, $x_2$, and $x_3$. The expressions for the energies and for the variables ${(N_1,N_2,N_3, \phi_{12}, \phi_{23})}$ of these critical points are provided in Table~\ref{tab2}. The sign of $J$ does not affect the results, so we fix $J>0$. For $\epsilon<0$ the minimum energy comes from $x_3$ and for $\epsilon>0$ it comes from $x_2$, as illustrated in Fig.~\ref{FigU0}~(a). 

The dependence of the occupation number of the wells on  $J/\epsilon$ for the stationary point with lowest energy is shown in Fig.~\ref{FigU0}~(b). When $|\epsilon| \gg |J|$,  the bosons are confined to well 1, which is understandable, since this is the deepest well. As the magnitude of the hopping amplitude gets larger than the magnitude of the tilt, the particles spread through the wells. In the extreme scenario of $|\epsilon|\ll |J|$ [not reached in Fig.~\ref{FigU0}~(b)], where the wells are nearly symmetric, half of the bosons become localized in well 2 and the other half splits equally between the wells 1 and 3 due to the symmetry between these wells.

\subsubsection{Quantum-classical correspondence: $U=0$}
\label{U0qu-cl} 

The results for the energy and for the mean values of the occupation numbers, $\langle \Psi_0 |\hat{N}_k |\Psi_0 \rangle$,  for the ground state $|\Psi_0 \rangle$ of the quantum Hamiltonian are also depicted in Figs.~\ref{FigU0}~(a)-(b) with circles. There is excellent agreement with the semiclassical results. In fact the agreement is perfect for even a single particle, because the model in the absence of interaction is in the semiclassical regime.  When $N=1$, there are just three eigenstates, all of them coherent, which can be verified from the perfect coincidence with the semiclassical values. For $N>1$, the quantum coherent states correspond to the minimal, intermediate, and maximal equilibrium points obtained in the semiclassical limit.

\subsection{Case $\pmb{J=0}$}

Following the definition in Ref.~\cite{Leggett2001}, the case $J=0$ is an extreme Fock regime, where quantum tunneling is forbidden and the occupation number (Fock) states are the eigenstates of the Hamiltonian. The solution of Eq.~(\ref{StabNiC}) for the semiclassical limit gives 5 different stationary points, $x_1$, $x_2$, $x_3$, $x_4$, and $x_5$, whose energies and variables ${(N_1,N_2,N_3, \phi_{12}, \phi_{23})}$ are given in Table~\ref{tab3}.

\begin{table*}
	\begin{center}
		\caption{{\bf Case $\pmb{J=0}$}: Exact expressions for the energies and variables ${(N_1,N_2,N_3, \phi_{12}, \phi_{23})}$ of the five stationary points. The points $x_2$ and $x_5$ exist only if $|U| \geq |\epsilon|/4$.}
		\begin{tabular}{|c|c|c|c|c|c|} \hline
			Stationary point & Energy $E/N$ & $N_1/N$ & $N_2/N$ & $N_3/N$ & $(\phi_{12},\phi_{23})$ \\ 
			\hline 
			$x_1$ & $U + \epsilon$ & 0 & 0 & 1 & arbitrary \\ 
			\hline 
	\rule{0pt}{\dimexpr.7\normalbaselineskip+2.5mm}
			$x_2$ & $-\dfrac{\epsilon^2}{16U}+\dfrac{\epsilon}{2}$ & 0 & $\dfrac{1}{2}+\dfrac{\epsilon}{8U}$ & $\dfrac{1}{2}-\dfrac{\epsilon}{8U}$ & arbitrary \\[0.2cm]  
			\hline 
			$x_3$ & $U$ & 0 & 1 & 0 & arbitrary\\ \hline $x_4$ & $U-\epsilon$ & 1 & 0 & 0 & arbitrary \\ 
			\hline 
	\rule{0pt}{\dimexpr.7\normalbaselineskip+2.5mm}
			$x_5$ & $-\dfrac{\epsilon^2}{16U}-\dfrac{\epsilon}{2}$ & $\dfrac{1}{2}+\dfrac{\epsilon}{8U}$ & $ \dfrac{1}{2}-\dfrac{\epsilon}{8U } $ & 0 & arbitrary \\[0.2cm] 
			\hline
		\end{tabular}    
		\label{tab3}
	\end{center}
\end{table*}

\begin{figure}[htb]
	\centering
	\includegraphics[width=1.\linewidth]{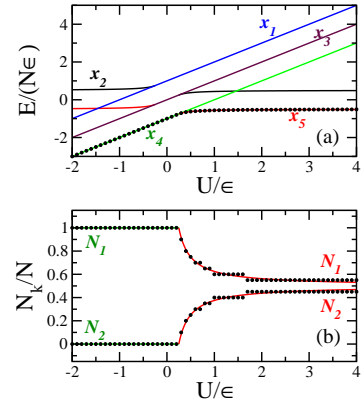}
	\caption{{\bf Case $\pmb{J=0}$}: Energies of the 5 stationary points (a) and the occupation numbers $N_1$ and $N_2$   for the minimum energy state (b);  $\epsilon>0$  and $N=20$.  Solid lines are for the semiclassical results and circles for the quantum ground state. $N_3$ is zero for the lowest energy state. }
	\label{FigJ0}
\end{figure}

The point $x_2$ ($x_5$) exists only if $|U| \geq |\epsilon|/4$, because according to Table~\ref{tab3}, for $|U|<|\epsilon|/4$, the occupation number $N_3$ ($N_2$) becomes negative. One also sees in Table~\ref{tab3} that changing the sign of $\epsilon$ simply interchanges  $x_1$ with $x_4$ and $x_2$ with $x_5$, so we assume that $\epsilon >0$. When $U<  \epsilon/4$, the minimum energy is determined by the stationary point $x_4$, while for $U> \epsilon/4$, the minimum energy is given by $x_5$. At $U= \epsilon/4$, there is a bifurcation, as shown in Fig.~\ref{FigJ0}~(a), which indicates a phase transition. The line at $U=\epsilon/4$ separates the plane $(U,\epsilon)$  in two different phases.

The phase transition is of second order. This can be verified by defining an arbitrary unitary vector, $\mathbf{v}=v_1\mathbf{x}+v_2\mathbf{y}$ and $\nabla=\left(\frac{\partial}{\partial U},\frac{\partial}{\partial \epsilon}\right)$, and using the directional derivative in the plane $(U,\epsilon)$. The first derivatives 
\[
\left. \left((\mathbf{v}\cdot\nabla)\frac{E_{4}}{N} \right) \right|_{U=\epsilon/4} =
 \left. \left((\mathbf{v}\cdot\nabla)\frac{E_{5}}{N} \right) \right|_{U=\epsilon/4}
=v_1-v_2,
\]
are equal, but not the second derivatives for $v_2\neq 4v_1$,
\begin{eqnarray}
\left. \left((\mathbf{v}\cdot\nabla)^2\frac{E_{4}}{N} \right) \right|_{U=\epsilon/4} &=&0 \nonumber \\
 \left. \left((\mathbf{v}\cdot\nabla)^2\frac{E_{5}}{N} \right) \right|_{U=\epsilon/4} 
&=&-\frac{\left(v_2-4v_1\right)^{2}}{8U}, \nonumber
\end{eqnarray}
which, according to the Ehrenfest criterion~\cite{Castanos2006}, implies a second-order phase transition.  

For the stationary point with lowest energy, $N_3=0$, so in Fig.~\ref{FigJ0}~(b) we show only $N_1$ and $N_2$ as a function of $U/\epsilon$. These two occupation numbers are good order parameters and clearly mark the phase transition at $U=\epsilon/4$. For $U<\epsilon/4$,  the lowest energy is obtained by having all bosons on well 1, while for $U>\epsilon/4$, the bosons distribute between well 1 and well 2, and they become equally distributed when $U\gg\epsilon$.  

\subsubsection{Quantum-classical correspondence: $J=0$}

Since the quantum Hamiltonian is diagonal in the Fock basis $|N_1,N_2,N_3\rangle$, finding the energy of the ground state comes down to finding the set of non-negative integers $(N_1,N_2,N_3)$ that minimizes the expression \begin{equation}
\frac{U}{N}\left(N_1-N_2+N_3\right)^2 + \epsilon\left(N_3-N_1\right) 
\end{equation} 
and satisfies $N_1+N_2+N_3=N$. As seen in Fig.~\ref{FigJ0}~(a), the quantum-classical correspondence for the lowest energy is excellent. In fact, for $U<\epsilon/4$, where all the bosons are on well 1,  the agreement holds for any number of bosons, including the limiting case $N=1$. When $U>\epsilon/4$, to lower the energy, the system needs to decrease the term $\frac{U}{N}\left(N_1-N_2+N_3\right)^2$, which can be achieved by placing part of the bosons on well 2. The quantum-classical agreement for the energy is good for small values of $N$ when the number of particles is even, $N=2, 4$, while for an odd number, larger $N$'s are needed.

In the case of the occupation numbers, since the mean values  $\langle \Psi_0 |\hat{N}_k |\Psi_0 \rangle$ are integers, when $U>\epsilon/4$, a large number of particles is needed for $\langle \Psi_0 |\hat{N}_1 |\Psi_0 \rangle$ and $\langle \Psi_0 |\hat{N}_2 |\Psi_0 \rangle$ to properly follow the classical curves, as shown in Fig.~\ref{FigJ0}~(b). The exception is the case where $U\gg \epsilon$ and $N_1=N_2$, for which the quantum-classical agreement holds for any even $N$.

\subsection{Case $\pmb{\epsilon=0}$}
\label{eps}

\begin{table*}
	\begin{center}
		\caption{{\bf Case $\pmb{\epsilon=0}$}: Exact expressions for the energies and variables ${(N_1,N_2,N_3, \phi_{12}, \phi_{23})}$ of the five stationary points. The points $x_4$ and $x_5$ exist only if $|U| \geq |J|/2$. Notice that $N_1=N_3$.}
		\begin{tabular}{|c|c|c|c|c|} 
		\hline
			Point & Energy $E/N$ & $N_1/N=N_3/N$ & $N_2/N$ & $(\phi_{12},\phi_{23})$ \\ 
			\hline 
			$x_1$ & $U$ & 1 & 0 & $(0,\pi)$  \\ \hline $x_2$ & $J$ & 1/4 & 1/2 & $(0,0)$ \\ 
			 \hline $x_3$ & $-J$ & 1/4 & 1/2 & $(\pi,\pi)$ \\ 
			 \hline $x_4$ & $U+\dfrac{J^{2}}{4U}$ & $\dfrac{1}{4}+\dfrac{1}{8}\sqrt{4-\dfrac{J^2}{U^2}}$ & $\dfrac{1}{2}-\dfrac{1}{4}\sqrt{4-\dfrac{J^2}{U^2}}$ & $(\pi,\pi)$ if $U/J<-1/2$, $(0,0)$ if $U/J>1/2$ \\ 
			 \hline $x_5$ & $U+\dfrac{J^{2}}{4U}$ & $\dfrac{1}{4}-\dfrac{1}{8}\sqrt{4-\dfrac{J^2}{U^2}}$ & $\dfrac{1}{2}+\dfrac{1}{4}\sqrt{4-\dfrac{J^2}{U^2}}$ & $(\pi,\pi)$ if $U/J<-1/2$, $(0,0)$ if $U/J>1/2$ \\ \hline
		\end{tabular}
		\label{tab1}
	\end{center}
\end{table*}

The solution of Eq.~(\ref{StabNiC}) for $\epsilon=0$ gives ten solutions for $(\rho_1,\rho_2,\rho_3)$ from which only five are different, $x_{1}$, $x_{2}$, $x_{3}$, $x_{4}$, and $x_{5}$. The energies and variables $(N_1,N_2,N_3,\phi_{12},\phi_{23})$ for these five stationary points are provided in Table~\ref{tab1}. Since the three wells have the same depth ($\epsilon=0$) and the Hamiltonian is invariant by switching wells 1 and 3, all the stationary points have $N_1=N_3$. Changing the sign of $J$ just interchanges $x_2$ with $x_3$, so we assume that $J>0$. The points $x_4$ and $x_5$ have the same energy and they only exist for $|U| \geq J/2$, when the $N_k$'s are real numbers.

\begin{figure}[htb]
	\centering
	\includegraphics[width=1.\linewidth]{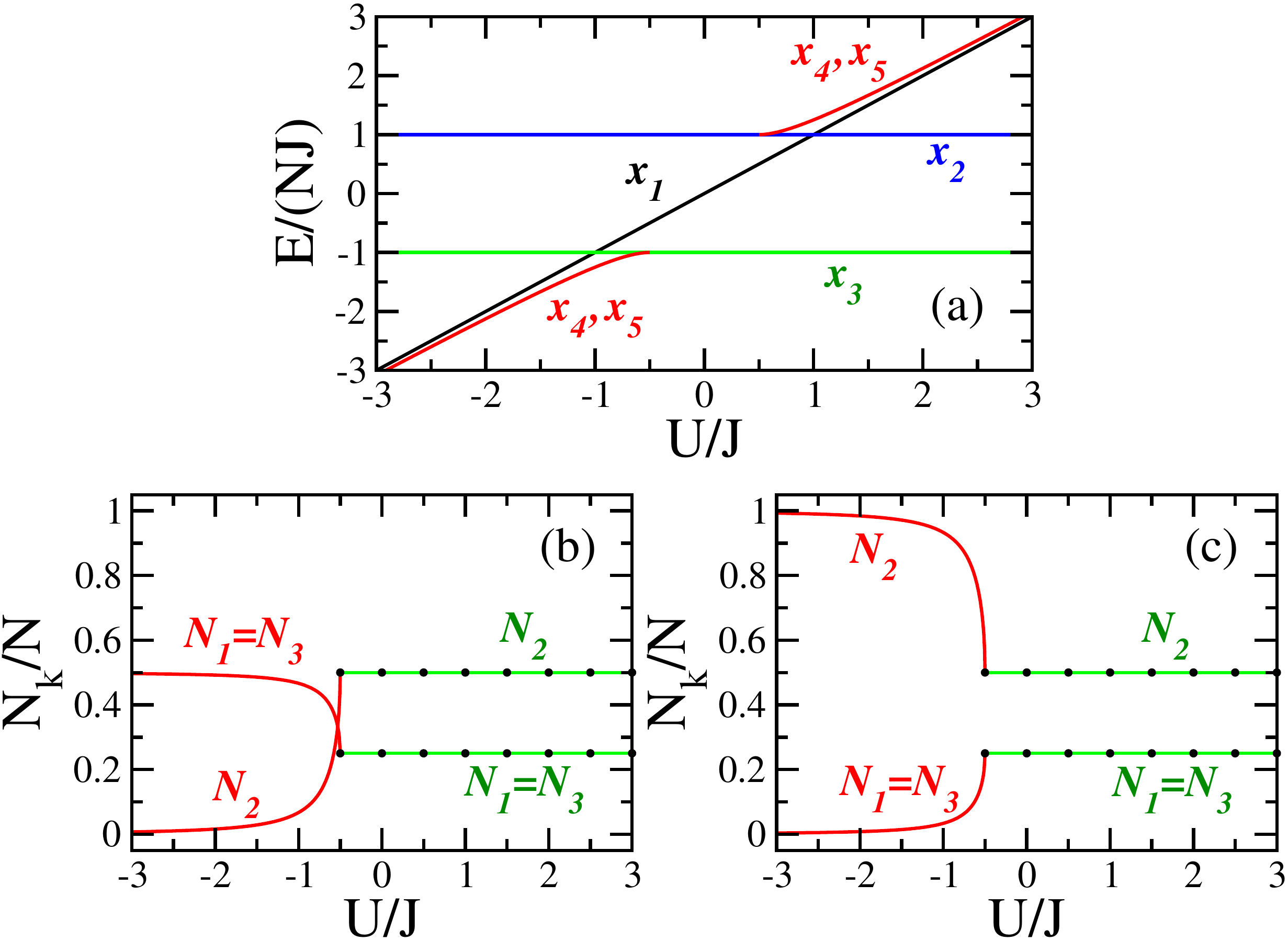}
	\caption{{\bf Case $\pmb{\epsilon=0}$}: Energies of the 5 stationary points (a) and the occupation numbers $N_k$  for the minimum energy state from  $x_4$ (b) and $x_5$ (c); $J>0$  and $N=20$.  Solid lines are for the semiclassical results and circles for the quantum ground state. The quantum results for $\langle N_k \rangle$ are shown only for $U>-J/2$. Results for $U<-J/2$ are discussed in Sec.~IV.
	}
		\label{FigE0}
	\end{figure}

In Fig.~\ref{FigE0}~(a), we show the energies for the five stationary points as a function of $U/J$. The minimum energy corresponds to the stationary points  $x_4$ and $x_5$ when $U<-J/2$, while for $U>-J/2$, the minimum energy  corresponds to $x_3$. Moving from positive $U$ towards the negative values, a bifurcation appears at  $U=-J/2$, which indicates a phase transition. Indeed, using  the directional derivatives in the plane $(U,J)$, we verify that this is a second-order phase transition. The line at $U=-J/2$ separates the plane $(U,J)$ in two phases.

The occupation numbers presented in Fig.~\ref{FigE0}~(b) for $x_4$ and in Fig.~\ref{FigE0}~(c) for $x_5$ behave as typical order parameters, exhibiting an abrupt change at the critical point $U=-J/2$. When the onsite interaction is attractive with $U< -J$, there are two possible scenarios: either the particles are equally distributed between wells 1 and 3 for $x_4$ [Fig.~\ref{FigE0}~(b)] or the particles are all contained in well 2 for $x_5$ [Fig.~\ref{FigE0}~(c)]. Approaching the phase transition, there is some leakage between those two scenarios. For $U>-J/2$, Fig.~\ref{FigE0}~(b) and Fig.~\ref{FigE0}~(c) become identical, since this is the region determined by $x_3$, where the bosons get spread out through the wells, half of them in well 2 and the other half equally distributed between wells 1 and 3.

\subsubsection{Quantum-classical correspondence: $\epsilon=0$}

\begin{figure}[htb]
	\centering
	\includegraphics[width=1.\linewidth]{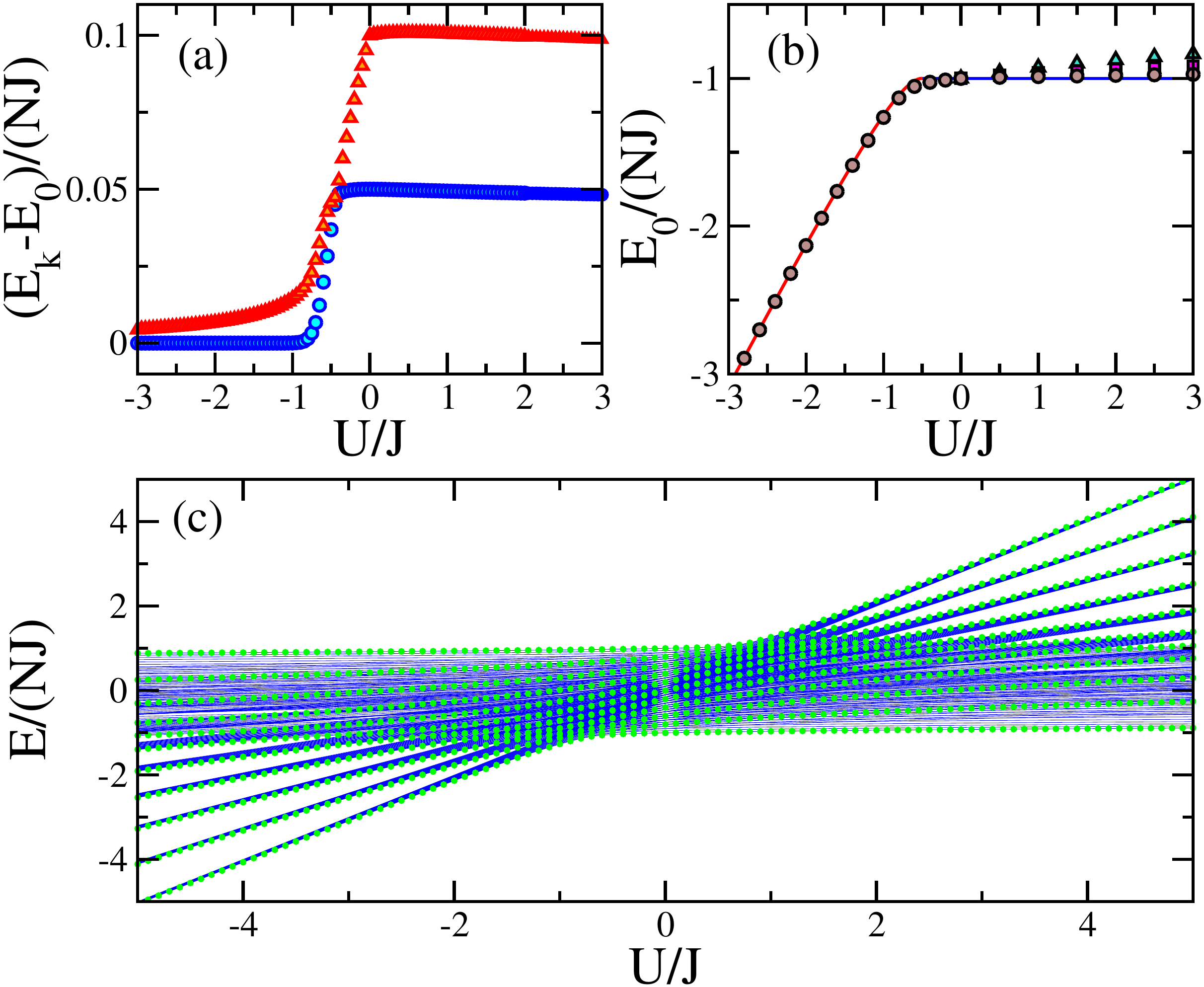}
	\caption{{\bf Case $\pmb{\epsilon=0}$}: Energy differences $E_1-E_0$ (circles) and $E_2-E_0$ (triangles)  in (a);  classical energy (solid line) and quantum ground state energy for $N=10$ (triangles), $N=20$ (squares), and $N=60$ (circles) in (b); energies of the triple-well system with $\epsilon=0$ (blue lines) and of the double well system (green circles). 
	In (a) and (c): $N= 20$.}
		\label{FigDeg}
	\end{figure}

The presence of two stationary points, $x_4$ and $x_5$, both with the same minimum energy for $U<-J/2$, gets manifested as a degenerate quantum ground state. This is shown in Fig.~\ref{FigDeg}~(a), where we depict the difference between the energies of the first excited state and the ground state as a function of $U/J$ (circles) and confirm that $E_1-E_0=0$ for $U<-J/2$. 

In Fig.~\ref{FigDeg}~(a), we also show  the energy difference between the second excited state and the ground state as a function of $U/J$ (triangles). The crossing point for the curves $(E_2 - E_0)/(NJ)$ and $(E_1 - E_0)/(NJ)$ indicates that $E_2 = E_1$, and since at this point the two energy differences are smaller than $10^{-1}$,  it can be seen that $E_2,E_1\sim E_0$. This reflects the semiclassical result at  $U/J=-1/2 $, where the green and red lines in Fig.~\ref{FigE0}~(a) coincide, that is, the stationary points $x_3$, $x_4$ and $x_5$ have the same energy. 

In Fig.~\ref{FigDeg}~(b), we compare the lowest classical energy (solid line) with the energy of the ground state (symbols). For  $U<-J/2$, the quantum-classical correspondence holds for few bosons, $N\gtrsim 5$, and for $U \ll -J/2$, the agreement is very good for as few as $N=2$ (not shown). In contrast, when the interaction becomes repulsive, $U/J>0$, a larger number of particles is needed for a good quantum-classical correspondence, as evident from the results for $N=10$ (triangles), $N=40$ (squares), and $N=60$ (circles) in Fig.~\ref{FigDeg}~(b).

When it comes to the comparison between the quantum and semiclassical results for the occupation numbers, the scenario changes. For $U/J \geq 0$, the agreement is already excellent for $N=1$ (not shown). This is understood, because for $U=0$, the system is in the semiclassical regime, as discussed in Sec.~\ref{U0qu-cl}, with $(N_1+N_3)/N = N_2/N = 1/2$; and for $U>J$,  the lowest energy is reached by vanishing the term $U(N_1-N_2+N_3)$, so the mean occupation numbers remain the same as those for $U=0$. In Fig.~\ref{FigE0}~(c), we show the mean values of the occupation numbers (circles) for $U>-J/2$ and $N=20$, and the quantum-classical agreement is indeed excellent.  For $U<-J/2$, due to the degeneracy of the ground state, any superposition of the packages centered at $| N_1^{x_4} N_2^{x_4} N_3^{x_4}  \rangle$ and at $| N_1^{x_5} N_2^{x_5} N_3^{x_5}  \rangle$ are valid ground states, so we leave the comparison between the semiclassical values of $N_k$'s and $\langle \Psi_0 |\hat{N}_k |\Psi_0 \rangle$ for this range of values of $U$ to the next section, where all parameters, including $\epsilon$, are non-zero, so the degeneracy is lifted.

It is informative to make a parallel between the symmetric triple-well potential and the symmetric double-well potential. By introducing a new annihilation operator ${\hat{b}=\sqrt{2}\hat{a}_1=\sqrt{2}\hat{a}_3}$, the quantum Hamiltonian (\ref{QH}) becomes equivalent to a double-well Hamiltonian, 
\begin{equation}
\label{two-well}
\hat{H}=\frac{U}{N}\left(\hat{b}^\dagger \hat{b} -\hat{N}_2\right)^2 + J\left(\hat{b}^\dagger \hat{a}_2+ \hat{a}_2^\dagger \hat{b} \right).
\end{equation}
Both Hamiltonians lead to a symmetry in the spectrum, where the eigenvalues changes sign, $E\rightarrow -E$,  as the interaction changes from attractive to repulsive, $U/J<0 \rightarrow U/J>0$. As $|U|/J$ increases from zero, the eigenvalues from $\hat{H}$ (\ref{QH}) [blue lines in Fig.~\ref{FigDeg}~(c)] approach those from $\hat{H}$ (\ref{two-well}) [green circles in Fig.~\ref{FigDeg}~(c)]. This becomes evident first for the extreme values, the ground state energy for $U/J<0$ and the highest eigenvalue for $U/J>0$, and it gradually reaches the other levels. In the limit $|U|/J \rightarrow \infty$, the Fock states become the eigenstates and the energies of both models then coincide. In the particular case of the triple-well system, it is clear that in this limit, the states $|N_1,N_2,N_3\rangle$ and $|N_2-n,N_1+N_3,n \rangle$ with $0 \leq n \leq N_2$ have the same energy and the degeneracy has order $N+2$ [or degeneracy of order $(N+2)/2$ for the state in the middle of the spectrum].

\section{Nonintegrable regime}

For the general case, where $U\neq 0$, $J \neq 0$, and $\epsilon \neq 0$, analytical solutions are no longer available, so our studies are numerical. As we explain in the appendix~\ref{App2}, to find $N_2$ it is necessary to solve a seventh degree polynomial, which in general does not have an analytical solution. 

\begin{figure}[t]
	\centering
	\includegraphics[width=1.\linewidth]{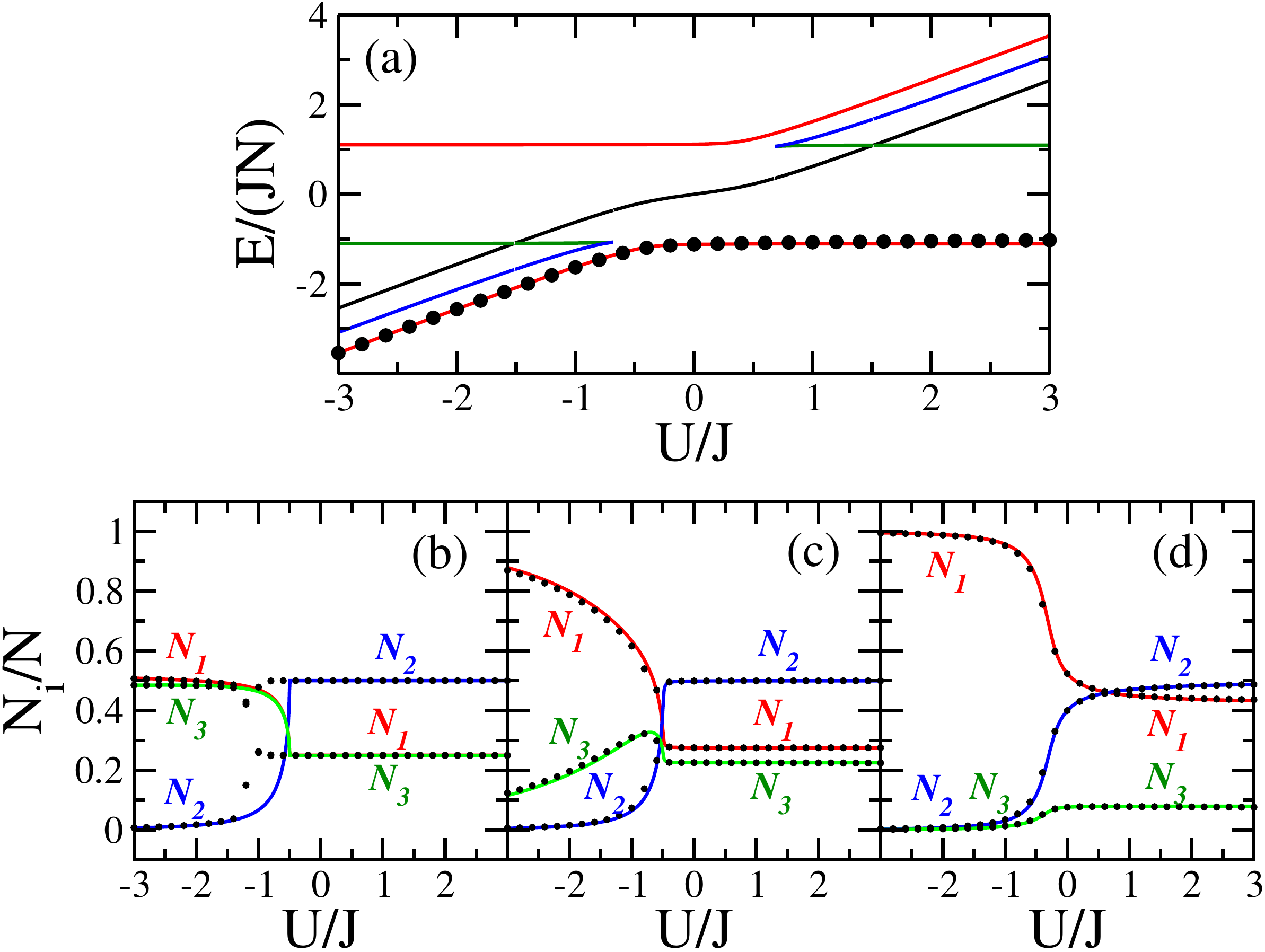}
	\caption{{\bf Case $\pmb{J, U, \epsilon \neq 0}$}: Energies of the 5 stationary points for $\epsilon/J =0.5$ (a) and the occupation numbers $N_k$ for the lowest energy for $\epsilon/J =10^{-3}$ (b), $\epsilon/J =0.05$ (c), and $\epsilon/J =0.5$ (d).   Solid lines are for the semiclassical results and circles for the quantum ground state;  $N=20$. Note: For the parameters used here, from the seven solutions of seventh degree polynomial  in the appendix~\ref{App2}, we find that only five are real.
	}
		\label{FigNonZero}
	\end{figure}

We start the study of this section in comparison with the one presented in Sec.~\ref{eps} for $\epsilon=0$. Here again we assume  that $J>0$ and also that $\epsilon >0$ ($\epsilon<0$ simply exchanges the roles of $N_1$ and $N_3$).  As seen in Fig.~\ref{FigNonZero}~(a), by tilting the potential, the bifurcation observed at $U/J=-1/2$  in  Fig.~\ref{FigE0}~(a) now vanishes. This implies that for $\epsilon \neq 0$, there is no longer a phase transition, but only crossovers. The crossovers can be seen in two directions as described in the next two paragraphs.

\begin{figure}[t]
	\centering
	\includegraphics[width=1.\linewidth]{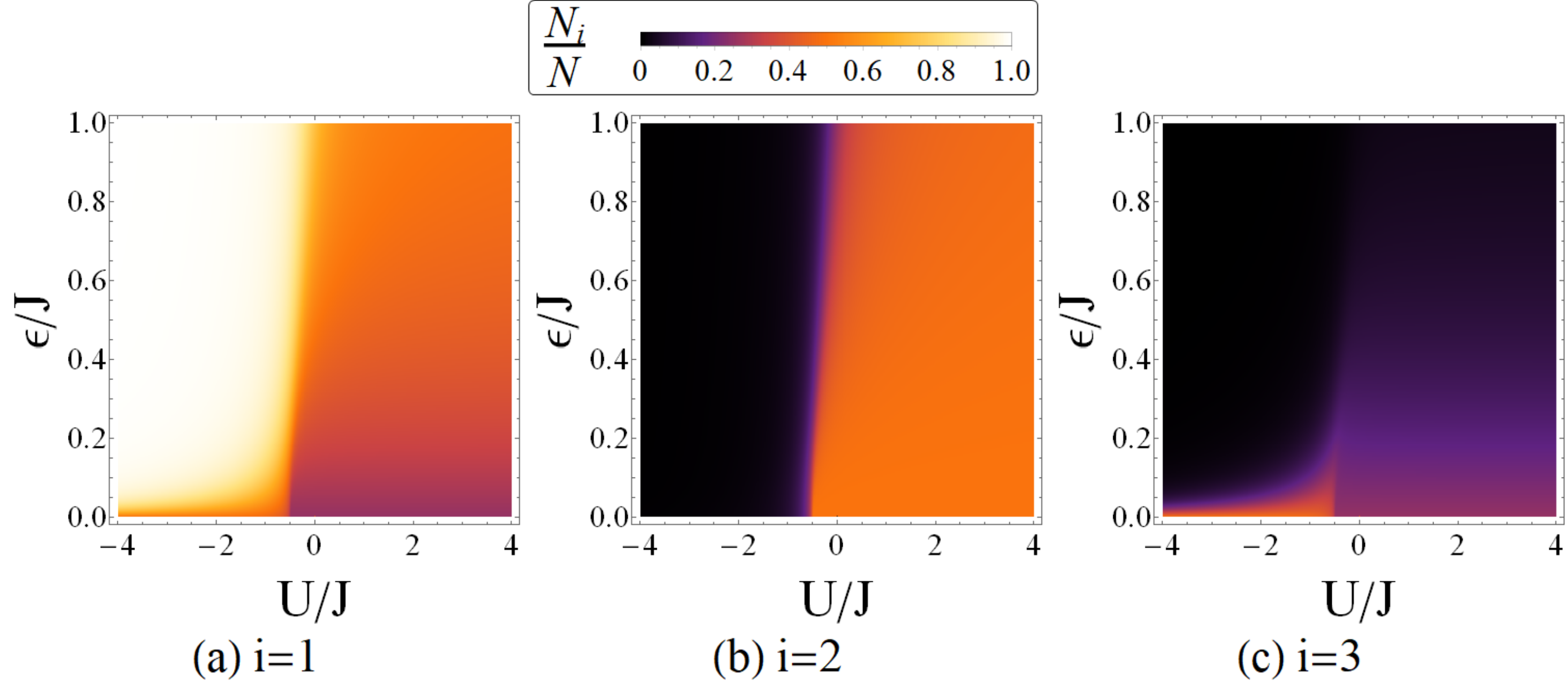}
	\caption{{\bf Case $\pmb{J, U, \epsilon \neq 0}$}: Density plots for the semiclassical results of the occupation numbers for different interaction strengths and tilt amplitudes.
	}
		\label{FigDensPlot}
	\end{figure}

For a fixed value of $\epsilon/J<1$, as $U/J$ goes from negative to positive values, there is a clear change in the behavior of the occupation numbers. For example, as seen in Figs.~\ref{FigNonZero}~(b)-(d) and also in the density plots in Fig.~\ref{FigDensPlot}~(b), $N_2/N$ is always zero when $U/J<-1/2$, but it becomes equal to $1/2$ for repulsive interaction. 

For a fixed value of $U/J$, as $\epsilon/J$ increases from zero, $N_3$ decays to zero and $N_1$ increases, as one sees by comparing Fig.~\ref{FigNonZero}~(b), Fig.~\ref{FigNonZero}~(c), and Fig.~\ref{FigNonZero}~(d) and by examining the density plots in Fig.~\ref{FigDensPlot}~(a) and Fig.~\ref{FigDensPlot}~(b). The change in the values of $N_3/N$ ($N_1/N$) from 1/2 to 0 (from 1/2 to 1) is abrupt when $U/J<-1/2$. This is evident from the very narrow lines seen close to $\epsilon/J=0$ for which $N_1\sim N_3 \sim 1/2$ in the density plots of Fig.~\ref{FigDensPlot}~(a) and Fig.~\ref{FigDensPlot}~(c). When the interaction is repulsive, $N_3/N$ decays smoothly  from $1/4$ to $0$ as $\epsilon/J$ increases and $N_1/N$ grows beyond 1/4 reaching a point, close to  $\epsilon/J \sim 1$, where $N_1=N_2=1/2$. If the tilt keeps increasing, so that $\epsilon/J>1$, $N_1$ naturally increases above $N_2$, but we disregard this trivial scenario and restrict our discussion to  $0<\epsilon/J<1$.

In the next two subsections, we have a closer look at the behaviors of the occupation numbers for  $U/J<-1/2$ and for repulsive interaction $U/J>0$.

\subsection{Degeneracy lift for $\pmb{U/J<-1/2}$}

The ground state degeneracy that the system had for $U/J<-1/2$ when $\epsilon=0$ is lifted for tiny values of the tilt. This has a dramatic effect on the values of the occupation numbers $N_1$ and $N_3$, as seen already for $\epsilon/J = 10^{-3}$ in Fig.~\ref{FigNonZero}~(b), where $N_1$ and $N_3$ no longer coincide. This high sensitivity of $N_1$ and $N_3$ to the amplitude of the tilt could be explored for developing new quantum sensors~\cite{Degen2017}. 

The idea of quantum sensing is to use quantum systems or quantum properties to measure physical quantities. Quantum sensors take advantage of the strong sensitivity of a quantum system to a perturbation. In the case of our model, keeping all other parameters constant, a dramatic change of the ground state indicates the presence of a tilt, and vice versa, a tiny variation of the tilt can be used to equate a state with the ground state. 

When $U/J<-1/2$, as $\epsilon/J$ increases [Figs.~\ref{FigNonZero}~(b)-(d)], the bosons migrate from well 3 to well 1, as expected, and well 2 remains empty. For values of the tilt as small as $\epsilon/J =0.3$, all particles are already trapped in well 1 and this picture does not change for larger values of $\epsilon/J$. The state becomes robust against changes in the value of the tilt.

\subsubsection{Quantum-classical correspondence: $\epsilon, J \neq 0$, $U/J<-1/2$}

In Sec.~\ref{eps}, where $\epsilon=0$, we mentioned that the quantum-classical correspondence for the lowest energy was excellent for few bosons when $U/J<-1/2$. For $\epsilon/J =0.5$, this agreement is already good for $N=2$. With respect to the occupation numbers, we postponed the discussion entirely to the present section, where the ground state is no longer degenerate. The general trend goes as follows. For a fixed value of the interaction strength with $U/J<-1/2$ and a fixed $N>1$, the agreement between the semiclassical values of $N_k$'s and $\langle \Psi_0 |\hat{N}_k |\Psi_0 \rangle$ improves as $\epsilon/J$ increases from zero, while for fixed values of $\epsilon/J$ and $N$, the agreement is better as $U/J \rightarrow -\infty$. In other words, in the vicinity of $U/J \sim -1/2$, a good quantum-classical correspondence requires a large number of particles, especially if $\epsilon/J$ is small and close to the critical point for $\epsilon=0$.

\subsection{Robustness of $\pmb{N_2}$ for $\pmb{U/J>0}$}

When the interaction is repulsive, the ground state for $0<\epsilon/J <0.1$ [Figs.~\ref{FigNonZero}~(b)-(c)] is a superposition of Fock basis very similar to that for $\epsilon =0$ [Fig.~\ref{FigE0}~(b)]. We indeed verified that the fidelity $|\langle \Psi_0 (\epsilon = 0) |\Psi_0 (\epsilon \neq 0) \rangle| >0.9$.  This implies that if one quenches the parameters $U/J$ and $\epsilon/J$ within a good range of values, the ground state remains invariant. The robustness of quantum states is a desirable feature in quantum information protocols where a specific property must be stable in spite of external perturbations.

As the tilt further increases, $\epsilon/J>0.1$, the fidelity decays significantly, because the particles from well 3 move to well 1 to keep the energy low. However, the occupation number of well 2 remains practically the same, $N_2/N \sim 1/2$,  up to $\epsilon/J \lesssim 1$. 

\subsubsection{Quantum-classical correspondence: $\epsilon, J \neq 0$, $U/J>0$}

With respect to the quantum-classical correspondence, it was shown in Sec.~\ref{eps} that for repulsive interaction and $\epsilon=0$, the agreement between the quantum and semiclassical results for the occupation numbers holds for a single boson. This agreement persists for $\epsilon/J<0.1$, but within a smaller range of values of the interaction, $0\leq U/J<1$.

\section{Discussion}

We studied a  system of interacting bosons in a triple-well potential, which can be experimentally realized with cold atoms. In two integrable limits of the model and using a semiclassical analysis, we identified two critical points associated with second-order quantum phase transitions. In the nonintegrable regime, the system exhibits crossovers under changes of the interaction strength or of the value of the potential tilt with respect to the tunneling amplitude. In all cases, we showed that the quantum and semiclassical results for the lowest energy and the corresponding occupation numbers of the three wells agree extremely well, and depending on  the Hamiltonian parameters and the quantity, the agreement may hold for even a single particle.

We found that in certain regions, the lowest energy state is either very sensitive or robust to changes in the Hamiltonian parameters, which could find application for quantum sensing or quantum information protocols, respectively. For attractive interaction, the occupation numbers of wells 1 and 3 are highly sensitive to the inclusion of the potential tilt. For repulsive interaction and small values of the tilt, the ground state is robust to changes in both the interaction strength and the tilt.

There are several interesting directions for extensions of the results presented in this work. They include the analysis of the higher levels in connection with the notion of excited state quantum phase transitions; the study of dynamics, in particular quench dynamics through different phases; and the addition of more wells~\cite{Grun2020,Grun2021}.

\begin{acknowledgements}
E.R.C. is grateful to the CBPF institution for the infrastructure offered during the writing process of this work and for the Brazilian agency CNPq (Conselho Nacional de Desenvolvimento Cient\'ifico e Tecnol\'ogico) for partial financial support. I.R. was also partially supported by CNPq. E.R.C., I.R. and L.F.S. thank Angela Foerster and Karin Wittmann Wilsmann for discussions. L.F.S. was supported by the NSF Grant no. DMR-1936006. J.G.H. was supported by DGAPA-UNAM project IN104020.
\end{acknowledgements}


\appendix

\section{Equations for the nonintegrable regime}
\label{App2} 

For the analysis of the equilibrium points in the nonintegrable regime, we resort again to Eq.~(\ref{StabNiC}). This set of four equations can be reduced to only two if we write $\rho_1$ and $\rho_3$ as functions of $\rho_2$ and $\lambda$, that is 

\begin{equation}\label{rho1}
\rho_1=\frac{\sqrt{2}}{2}\frac{J\rho_2}{\lambda+\epsilon-2U\left(1-2\rho_2^2\right)} ,
\end{equation}   

\begin{equation}\label{rho3}
\rho_3=\frac{\sqrt{2}}{2}\frac{J\rho_2}{\lambda-\epsilon-2U\left(1-2\rho_2^2\right)} .
\end{equation} 
The two equations that determine $\rho_2$ and $\lambda$ become
\begin{align}\label{X1}
	\frac{1}{\rho_2^2}-1=&\frac{J^2}{2}\left[\frac{1}{\left[\lambda+\epsilon-2U\left(1-2\rho_2^2\right)\right]^{2}}\right. \nonumber \\ &\left.\,\,\,\,\,\,\,\,\,\,+\frac{1}{\left[\lambda-\epsilon-2U\left(1-2\rho_2^2\right)\right]^{2}}\right],
\end{align}	 

\begin{align}\label{X2}
	L+2U\left(1-2\rho_2^2\right)=&\frac{J^2}{2}\left[\frac{1}{\lambda+\epsilon-2U\left(1-2\rho_2^2\right)}\right. \nonumber \\ &\left.\,\,\,\,\,\,\,\,\,\,+\frac{1}{\lambda-\epsilon-2U\left(1-2\rho_2^2\right)}\right].
\end{align}

Using the definition 
\begin{equation}
\label{L}
X \equiv \lambda-2U(1-2\rho_2^2),
\end{equation} 
we can reduce Eq.~(\ref{X1}) to one quadratic polynomial in $X^2$ whose roots determine the expression $X(\rho_2)$. In particular,
\begin{equation}\label{X}
X=\pm\sqrt{\epsilon^2+\frac{J\rho_2^2}{2(1-\rho_2^2)}\left[J\pm\sqrt{J^2+8\left(\frac{1-\rho_2^2}{\rho_2^2}\right)\epsilon^2}\right]} .
\end{equation} 
For each $\rho_2$, there are 4 possible values of $X$, of which only one gives the correct result. The correct result corresponds to the value of $\lambda$ in Eq.~(\ref{L}) that satisfies the four equations in Eq.~(\ref{StabNiC}). 

A better option is to eliminate $\lambda$ and reduce Eq.~(\ref{X1}) and Eq.~(\ref{X2}) to a seventh degree polynomial in $\rho_2^2$ whose roots are the equilibrium values of $\rho_2$. Explicitly,
\begin{equation}\label{Poly}
\sum_{m=0}^{7}C_m(U^2,J^2,\epsilon^2)(\rho_2^2)^m=0.
\end{equation}
The coefficients of this polynomial are shown below,
\begin{equation}
C_0=-\epsilon^2 J^4 ,
\end{equation}
\begin{align}
C_1=4 \epsilon^4 J^2 + 5 \epsilon^2 J^4  + J^6 + 64 \epsilon^2 J^2  U^2	,
\end{align}	
\begin{align}
	C_2=&-4 \epsilon^6  - 12 \epsilon^4 J^2  - 12 \epsilon^2 J^4  - 4 J^6 \nonumber\\ &+ 
	128 \epsilon^4  U^2 - 576 \epsilon^2 J^2  U^2 - 16 J^4 U^2\nonumber \\& - 
	1024 \epsilon^2 U^4	,
\end{align}	
\begin{align}
	C_3=&4 \epsilon^6 + 12 \epsilon^4 J^2 + 12 \epsilon^2 J^4 + 4 J^6 - 640 \epsilon^4 U^2 \nonumber \\ &+ 
	1856 \epsilon^2 J^2 U^2 + 80 J^4 U^2 + 9216 \epsilon^2 U^4	,
\end{align}	
\begin{align}
	C_4=&1024 \epsilon^4 U^2 - 2560 \epsilon^2 J^2 U^2 - 128 J^4 U^2\nonumber \\ &- 
	32768 \epsilon^2 U^4	,
\end{align}	
\begin{align}
	C_5=&-512 \epsilon^4 U^2 + 1280 \epsilon^2 J^2 U^2 + 64 J^4 U^2\nonumber \\ &+ 57344 \epsilon^2 U^4	,
\end{align}	
\begin{equation}
C_6=-49152 \epsilon^2 U^4 ,
\end{equation}

\begin{equation}
C_7=16384 \epsilon^2 U^4 .
\end{equation}
As the roots of a seventh-degree polynomial do not have, in general, analytical expressions, we resort to numerical calculations. Taking only the positive solutions for $\rho_2$, we have at most seven real solutions for each value of $(U,J,\epsilon)$. With Eq.~(\ref{rho1}) and  Eq.~(\ref{rho3}), we can calculate the coordinates $(N_1,N_2,N_3)$ and the phases differences (depending on the sign of $\rho_1$ and $\rho_3$), and using Eq.~(\ref{CH1EqC}), we get the energy per particle for each stationary point.

\nocite{*}

\begin{thebibliography}{51}
\bibitem[A]{fn1}%
 Codes and data shall be provided upon request.
\end{thebibliography}

%

\end{document}